\documentclass[amsmath,amssymb, 12pt,a4paper, tightenlines]{revtex4-1}

\usepackage[usenames,dvipsnames,svgnames,table]{xcolor}

\usepackage{graphicx}
\usepackage{epstopdf}
\usepackage{amssymb,amsfonts,amsmath}
\usepackage{siunitx}
\usepackage{bm}
\usepackage{cleveref}

\usepackage{color}
\definecolor{blue}{HTML}{377EB8}
\definecolor{red}{HTML}{E31A1C}

\begin{document}

\newcommand\solidrule[1][1cm]{\rule[0.5ex]{#1}{1.25pt}}
\newcommand\dashedrule{\mbox{%
  \solidrule[.6mm]\hspace{.5mm}\solidrule[.6mm]\hspace{.5mm}\solidrule[.6mm]}}

\title{Undulatory swimming in shear-thinning fluids:\\Experiments with \emph{C. elegans}}

\author{David~A.~Gagnon}
\author{Nathan~C.~Keim}
\author{Paulo~E.~Arratia}
\email{parratia@seas.upenn.edu}
\affiliation{Department of Mechanical Engineering and Applied Mechanics, University~of~Pennsylvania, Philadelphia,~PA 19104}


\date{\today}

\begin{abstract}
The swimming behaviour of microorganisms can be strongly influenced by the rheology of their fluid environment. In this manuscript, we experimentally investigate the effects of shear-thinning viscosity on the swimming behaviour of an undulatory swimmer, the nematode \textit{Caenorhabditis elegans}. Tracking methods are used to measure the swimmer's kinematic data (including propulsion speed) and velocity fields. We find that shear-thinning viscosity modifies the velocity fields produced by the swimming nematode but does not modify the nematode's speed and beating kinematics. Velocimetry data show significant enhancement in local vorticity and circulation, and an increase in fluid velocity near the nematode's tail, compared to Newtonian fluids of similar effective viscosity. These findings are compared to recent theoretical and numerical results.
\end{abstract}

\maketitle

\section{Introduction}
Swimming organisms are an essential component of numerous biological processes, from reproduction in mammals~\citep{Fauci2006} and infections such as Lyme disease~\citep{Harman2012} to bio-degredation in soil~\citep{Alexander1991}. Examples of swimming organisms include sperm cells~\citep{Fauci2006}, bacteria and many types of protozoa~\citep{Lauga2012}, and multi-cellular organisms such as nematodes~\citep{Sznitman2010PoF}. Due to their small length scales ($L\lesssim 1$~mm), many of the above mentioned organisms live in a low Reynolds-number regime ($Re\ll1$) where linear viscous forces are much larger than nonlinear forces from inertia. At low $Re$, locomotion requires non-reciprocal deformations to break time-reversal symmetry, known as the ``scallop theorem''~\citep{Purcell1977}.

Recently, there has been much interest in understanding the swimming behaviour of microorganisms at low $Re$ in simple, Newtonian fluids~\citep{Lauga2009, Guasto2010, Saintillan2012, Lauga2012}. However, many organisms swim in non-Newtonian environments such as mucus, blood, and soil~\citep{Harman2012, Alexander1991}. An important feature of non-Newtonian fluids is that they often exhibit viscoelasticity and shear-rate dependent viscosity. While much work has been devoted to the effects of fluid elasticity on the swimming of microorganisms~\citep{Lauga2007, Fu2009, Fu2010, Teran2010, Shen2011, Liu2011}, there are relatively few studies of swimming in shear-thinning fluids.

To date, major studies of the effects of shear-thinning viscosity have been theoretical~\citep{Velez2013} and numerical~\citep{MJ2012, MJ2013}. The theoretical analysis focused on a two-dimensional, infinite waving sheet immersed in a model Carreau (shear-thinning) fluid, and found a non-Newtonian contribution to propulsion speed to fourth order in amplitude when the sheet was extensible, but no non-Newtonian contribution to propulsion speed when an inextensible condition was applied~\citep{Velez2013}. Additionally, this analysis suggested the cost of transport was reduced and the flow field was modified, with increased vorticity near the sheet. Additionally, for a finite swimmer, this analysis suggested the cost of transport was reduced and the flow field was modified, with increased vorticity near the sheet~\citep{Velez2013}. Separate simulations~\citep{MJ2012, MJ2013} also using the Carreau model suggested that undulatory swimmers with a head or ``payload'' (similar to a sperm cell) are assisted by shear-thinning viscosity, resulting in increased speed and that the swimmer's motion results in an envelope of thinned fluid around the body. 

Despite these recent and important efforts, there is still a dearth of experimental investigations of swimming in shear-thinning fluids, and the effects of rate-dependent viscosity on swimming remain unclear. Experiments with a mechanical model system~\citep{Dasgupta2013} finds a decrease in propulsion for fluids possessing both shear-thinning and viscoelastic behavior, while the swimming speed of \textit{C. elegans} is shown to be unaffected by shear-thinning viscosity~\citep{Shen2011} although only a single data point is available. Here, we experimentally investigate the effects of shear-thinning viscosity on the swimming behaviour of a model biological organism, the nematode \textit{Caenorhabditis elegans}.  The nematode's position and swimming stroke is tracked using in-house software~\citep{RSznitman2010} and flow fields are obtained using particle tracking methods~\citep{Sznitman2010PoF}. Results show there is no change in the nematode's kinematics due to shear-thinning effects. Yet with this unchanged swimming stroke, the nematode in shear-thinning fluid generates a remarkably different flow field, with enhanced vorticity and an altered spatial pattern of fluid velocity. We compare these experimental results to recent analysis~\citep{Velez2013} and numerical simulations~\citep{MJ2012, MJ2013}.

\begin{figure}
  \centerline{\includegraphics[width=11.5cm]{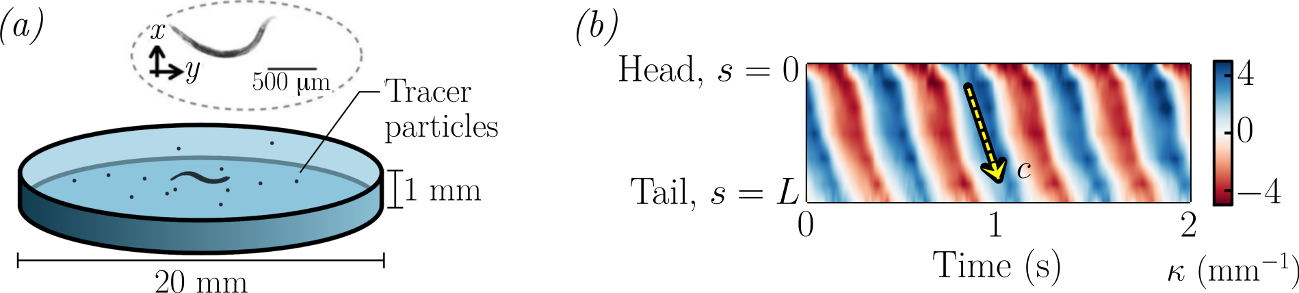}}
  \caption{(Colour available online) \textit{(a)} Schematic of nematode \textit{C.\ elegans} swimming in a sealed fluidic chamber. Inset: sample image. \textit{(b)} Nematode's body curvature during swimming. Curvature plot for approximately four beating cycles illustrates characteristic travelling wave propagating from head to tail. Wave speed $c$ is highlighted by the yellow dashed arrow.}
\label{fig1}
\end{figure}

\section{Experimental Methods}
The swimming behaviour of the nematode \textit{C.\ elegans} in shear-thinning fluids is investigated in a sealed acrylic chamber that is 2~cm in diameter and 1~mm in depth (figure~\ref{fig1}\textit{(a)}) using a microscope and a high-speed camera. \textit{C.\ elegans} is a roundworm widely used for biological research \citep{Rankin2002} that swims by generating travelling waves~\citep{Sznitman2010PoF}; organisms are approximately 1~mm in length and 80~\si{\micro}m in diameter. Two main types of experiments are performed: \textit{(i)} tracer particle velocimetry, used to obtain velocity fields, and \textit{(ii)} nematode tracking, used to obtain kinematic data such as swimming speed, beating frequency and amplitude. Images are recorded with the focal plane at the centre of the chamber to avoid movies with nematode-wall interactions; out-of-plane recordings are discarded. Kinematics data are an average of at least 15 recordings in each fluid tested. 

The nematode's swimming kinematics are obtained from videos using in-house software~\citep{RSznitman2010}. The software extracts the nematode's centroid position and body shape-line, and computes quantities such as swimming speed $U$ and body curvature $\kappa$. Swimming speed is obtained by differentiating the nematode's centroid position with time, and we define the positive $y$-axis as the swimming direction (see figure~\ref{fig1}\textit{(a)}). The body curvature is defined as $\kappa = \delta \phi / \delta s$, where $\phi$ is the angle between a fixed reference axis and the tangent to the body shape-line, at each point~$s$ along the body contour, where $s$ is an arc length parameterisation. Figure~\ref{fig1}\textit{(b)} shows the evolution of $\kappa(s,t)$ for approximately four beating cycles, revealing periodic lines that propagate in time from head ($s=0$) to tail ($s=L$),  illustrating the characteristic travelling wave of undulatory swimming. The slope and rate of occurrence of these lines represent wave speed~$c$ (figure~\ref{fig1}\textit{(b)}, yellow dashed line) and beating frequency~$f$, respectively. In water-like buffer solutions, we find $U\approx0.35$ mm/s, $f\approx2$~Hz, and $c\approx5$~mm/s. The Reynolds number, defined as $Re = \rho U L / \eta$, is approximately 0.35, where $L$ is the nematode's length (1~mm), $\rho$ is the fluid's density ($10^3$~kg/m$^3$), and $\eta$ is the fluid's viscosity (1~mPa$\cdot$s), indicating that viscous forces dominate the flow. The range of Reynolds numbers for all experiments is $10^{-4} < Re \leq 0.35$.

Newtonian fluids of various viscosities are: a water-like buffer solution (M9 salt solution~\citep{Brenner1974}, $\eta=1$~mPa$\cdot$s); very dilute solutions of the polymer carboxymethyl cellulose (CMC, $7\times10^5$ MW, Sigma Aldrich 419338) in M9 salt solution ($2 \leq \eta \leq 12$~mPa$\cdot$s); and mixtures of halocarbon oils (Sigma Aldrich H8773 and H8898, $27 \leq \eta \leq 700$~mPa$\cdot$s ). From among these fluids, we can obtain (Newtonian) viscosities ranging from 1~mPa$\cdot$s to 700~mPa$\cdot$s. We note that while aqueous solutions of CMC can exhibit viscoelasticity (CMC is a flexible polymer), one can drastically minimize the elasticity of such solutions by using low polymer concentrations and salt in solution. Under those conditions (low CMC concentration and presence of salt) CMC solutions exhibit Newtonian viscosity behavior and negligible elasticity.  

\begin{figure}
\centerline{\includegraphics[width=9.0cm]{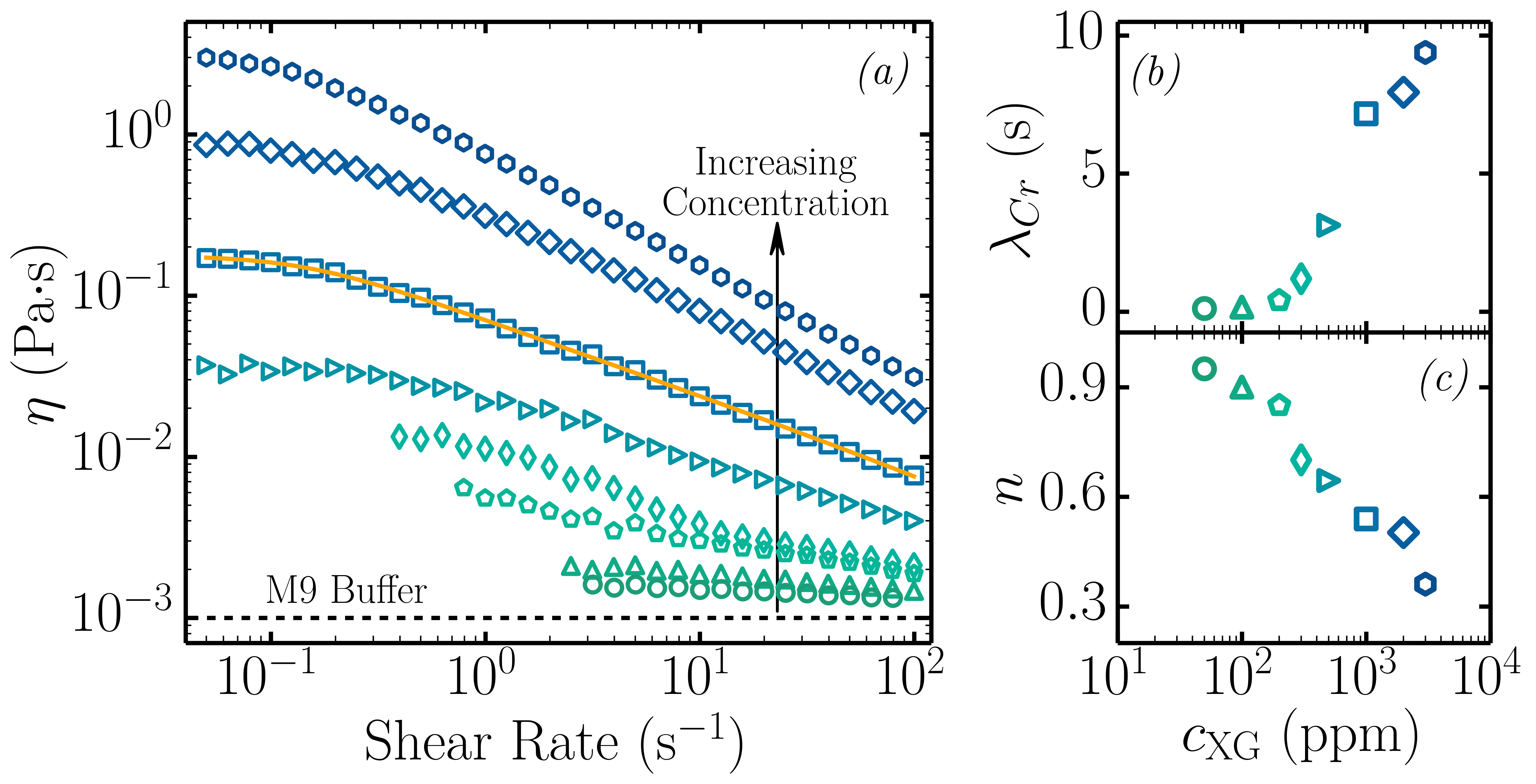}}
\caption{(Colour available online) \textit{(a)}~Shear viscosity of mixtures of xanthan gum (XG) in M9 buffer solution from 50~ppm to 3000~ppm, which show behaviour characteristic of Carreau-Yasuda model fluids. Dashed line indicates the viscosity of the solvent (M9). Orange line shows a sample fit to the Carreau model. \textit{(b)}~Carreau timescale~$\lambda_{Cr}$ and \textit{(c)}~shear-thinning index~$n$ calculated using least-squares fit to shear viscosity data (see fit line in \textit{(a)}).}
\label{fig2}
\end{figure}

Shear-thinning (ST) fluids are dilute aqueous solutions of the stiff ``rod-like" polymer xanthan gum in M9 buffer (XG, $2.7 \times 10^{6}$ MW, Sigma Aldrich G1253). XG solutions are known to be shear-thinning, with negligible elasticity~\citep{Wyatt2009,Shen2011}. We control the strength of shear-thinning by varying XG concentration in solution from 50~ppm to 3000~ppm. Fluids are characterized using a stress-controlled cone-plate rheometer. Figure~\ref{fig2}\textit{(a)} shows shear viscosity as a function of shear rate for all XG solutions; Newtonian fluids are not shown. Strong ST viscosity is found for the highest-concentration solutions (1000, 2000, and 3000 ppm) while nearly water-like viscosity is found for the lowest-concentration solutions (50 and 100~ppm). The viscosity data is fit to the Carreau-Yasuda model~\citep{Carreau1997}, defined as $\eta (\dot{\gamma})=\eta_{\infty} + (\eta_0 - \eta_{\infty}) (1 + (\lambda_{Cr} \dot{\gamma})^2)^{\frac{n-1}{2}}$, where $\eta(\dot{\gamma})$ is the fluid's shear-rate-dependent viscosity, $\eta_0$ is the zero-shear viscosity, $\eta_{\infty}$ is the infinite-shear viscosity, and $n$ is the power-law index. The characteristic timescale $\lambda_{Cr}$ is the inverse of the shear rate at which the fluid transitions from Newtonian-like to power-law behaviour; it is also the timescale for ST effects to develop or fade away when $\dot \gamma$ is changed~\citep{Carreau1997}.

Figures~\ref{fig2}\textit{(b)} and~\ref{fig2}\textit{(c)} show $\lambda_{Cr}$ and $n$ as a function of polymer concentration $c_{\mathrm{XG}}$. As $c_{\mathrm{XG}}$ increases, the rise in $\lambda_{Cr}$ indicates that the fluid requires a lower shear rate to access the shear-thinning regime. The power-law index~$n$ simultaneously decreases, indicating that the fluid becomes more shear-thinning. Note that the power in the Carreau-Yasuda model is defined as $(n-1)/2$; hence a decrease in $n$ means an increase in shear-thinning behaviour. The Carreau number $Cr = \lambda_{Cr} \dot{\gamma}$, indicates the degree to which a flow is shear-thinning: if $Cr < 1$, the viscosity is Newtonian-like; if $Cr > 1$, the viscosity is shear-thinning~\citep{Carreau1997}.

\section{Swimming Kinematics}
\begin{figure}
  \centerline{\includegraphics[width=11.5cm]{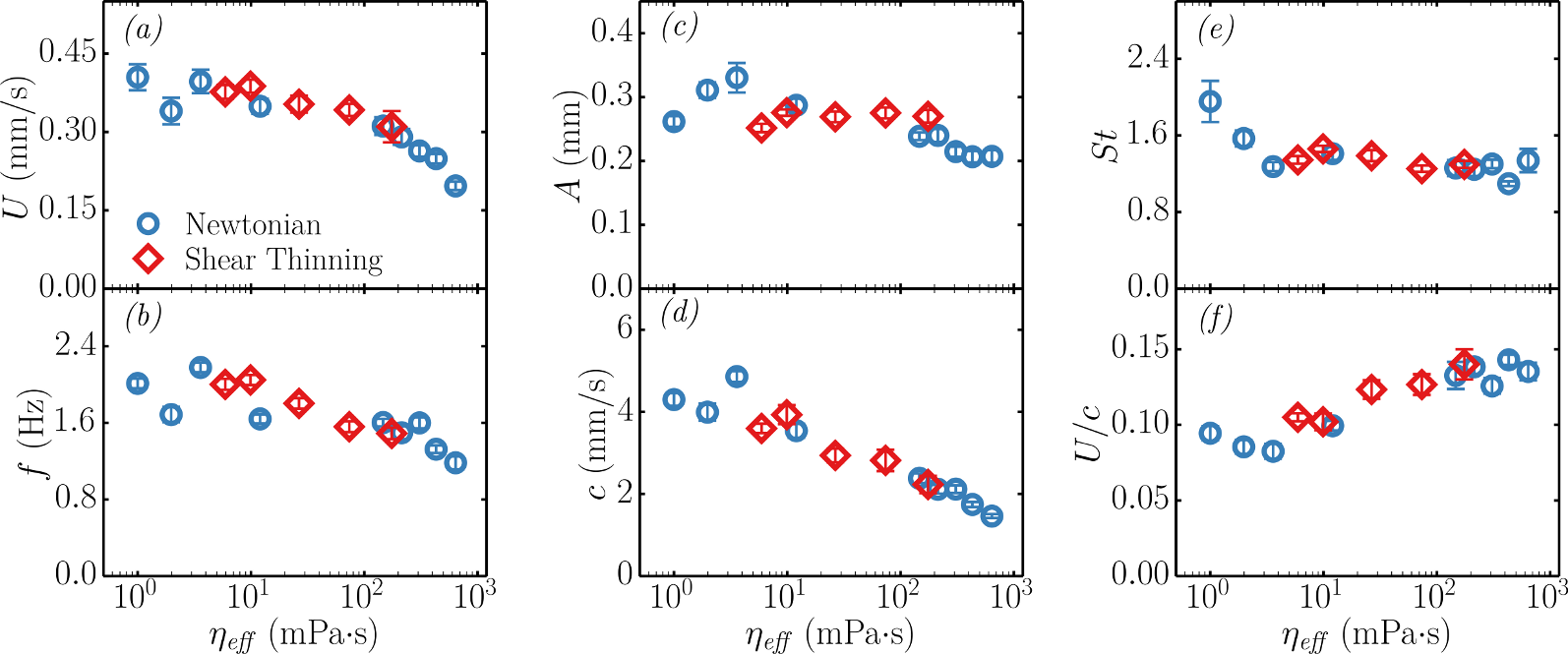}}
  \caption{(Colour available online) Summary of nematode kinematics for Newtonian~({\color{blue}$\bm{\circ}$}) and shear-thinning~({\color{red}$\bm{\diamond$}}) fluids. \textit{(a)}~Swimming speed~$U$, \textit{(b)}~frequency~$f$, \textit{(c)}~head amplitude~$A$, \textit{(d)}~wave speed~$c$, \textit{(e)}~Strouhal number~$St$ and \textit{(f)}~kinematic efficiency~$U/c$ as a function of effective viscosity $\eta_{\mathit{eff}}$. Each data point represents the mean and standard error of approximately 15 recordings.}
\label{fig3}
\end{figure}

We begin by investigating the swimming kinematics of nematodes in Newtonian and shear-thinning (ST) fluids. Figure~\ref{fig3}\textit{(a-f)} show the nematode's \textit{(a)} swimming speed~$U$, \textit{(b)} beating frequency~$f$, \textit{(c)} beating amplitude~$A$, \textit{(d)} wave speed~$c$, \textit{(e)} Strouhal number~$St=fA/U$, and \textit{(f)} kinematic efficiency~$U/c$ as a function of effective viscosity~$\eta_{\mathit{eff}}$ for both Newtonian and ST fluids. Note that for ST fluids, $\eta_{\mathit{eff}}$ is defined as the average viscosity over a characteristic range of shear rates, $U/L \leq \dot{\gamma} \leq 2fA/d$. The rate $U/L \sim 0.35$~s$^{-1}$ describes the forward progress of the entire nematode, and the rate $2fA/d \sim 15$~s$^{-1}$ describes the nematode's local transverse motion, where $d/2\sim 40$~$\si{\micro}$m is the body radius. These estimates are in good agreement with measured shear rates.

The data in figure~\ref{fig3} show that the nematode's swimming kinematics in ST fluids are very similar to those in Newtonian fluids of the same effective viscosity. This suggests that the nematode's kinematics are responding solely to bulk viscous effects and not local shear-thinning effects. For example, the nematode's swimming speed~$U$ (figure~\ref{fig3}\textit{(a)}) is approximately constant for both ST and Newtonian fluids up to $\eta_{\mathit{eff}} \approx 50$~mPa$\cdot$s. A decrease in $U$ is also found for both ST and Newtonian fluids for $\eta_{\mathit{eff}} > 50$~mPa$\cdot$s, which is in agreement with the power-limited nature of \textit{C.\ elegans}~\citep{Shen2011}. 

Overall, we find no evidence that shear-rate-dependent viscosity influences the swimming stroke or speed of \textit{C.\ elegans}. Our findings seem to be in agreement, at least in part, with a recent theoretical work~\citep{Velez2013}, in which the authors find no ST-induced changes to swimming speed for an inextensible sheet. For an extensible sheet, the same study finds an increase in swimming speed in shear-thinning fluids. We note, however, that there are differences between our experimental system and the aforementioned theoretical study. For example, the nematode is of finite-length and swims using large-amplitude waves that decay from head to tail, while the theoretical work focuses on an infinite, two-dimensional waving sheet of prescribed, small-amplitude kinematics. Thus, quantitative agreement is not expected.

This theoretical work also proposed a reduction in the cost of transport in a ST fluid. At low $Re$, power scales as $P \sim \eta V^2$; for swimmers with identical kinematics (same $V$ near the body) and similar effective shear viscosities $\eta_{eff}$, we expect $P$ and therefore the cost of transport to be similar. We note that this is just an approximation and a more rigorous calculation using spatially resolved velocity fields (near the nematode's body) and a constitutive model for the ST fluid is necessary~\citep{Sznitman2010PoF}.

We can also compare our findings to numerical simulations~\citep{MJ2012, MJ2013} that have predicted an enhancement in swimming speed; however, the enhancement predicted by this study is for a swimmer with an elliptical head and a linearly increasing amplitude from head to tail; since \textit{C. elegans} has no ``head'' and conversely has a decreasing amplitude from head to tail, these differences in geometry and stroke may explain the discrepancy between these simulations' predictions and our experimental measurements~\citep{MJ2012, MJ2013}. 

\section{Velocimetry, Flow Fields, and Streamlines}
In the preceding section, we showed that the swimming strokes of nematodes in ST and Newtonian fluids are quite similar (if not identical). However, the same kinematics may generate different velocity fields. Since the swimming stroke corresponds to the imposed fluid boundary conditions, in Newtonian fluids we expect flow fields to have a common morphology, with velocity magnitudes merely scaling with viscosity. However, under the same boundary conditions, a non-Newtonian constitutive relation might result in measurable changes to the flow fields. To investigate this possibility, we use particle tracking to measure the flow fields around swimming \textit{C.\ elegans} in ST and Newtonian fluids. We seed the fluids with $3.1$~\si{\micro}m-diameter fluorescent polystyrene spheres, which are tracked in space and time~\citep{Sznitman2010PoF, Gagnon2013}. The particles are very dilute ($<0.5$\% by volume) and do not alter the fluid rheology.  

\begin{figure}
  \centerline{\includegraphics[width=9.cm]{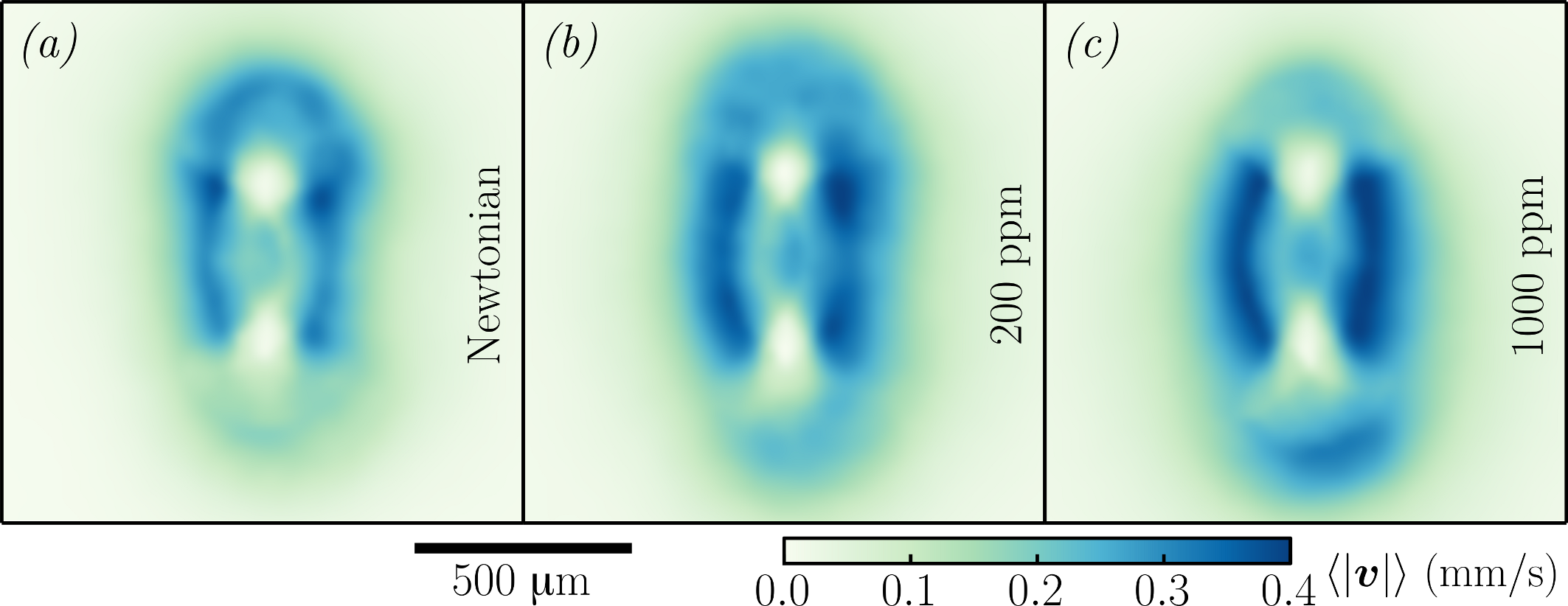}}
\caption{(Colour available online) \textit{(a-c)} Time-averaged fluid velocity magnitude for one beating cycle of \textit{C.\ elegans} in \textit{(a)} Newtonian buffer solution, \textit{(b)} shear-thinning 200 ppm XG solution, and \textit{(c)} shear-thinning 1000 ppm XG solution.}
\label{fig4}
\end{figure}

Because the nematode's swimming gait is periodic (Fig~\ref{fig1}\textit{(b)}), we can significantly increase the spatial resolution of our flow field measurements by using a phase averaging scheme. We condense multiple beating cycles into a single ``master'' cycle by matching frames with the same curvature profile $\kappa(s,t)$, using a least-squares algorithm. This master cycle comprises body shapes and velocity fields at approximately 65 different phases. For more details on this technique, see~\citet{Sznitman2010PoF}. 

Figure~\ref{fig4} shows time-averaged velocity magnitude fields in \textit{(a)}~buffer solution (Newtonian), \textit{(b)}~200~ppm XG solution, and \textit{(c)}~1000~ppm XG solution over one beating cycle. Two features are immediately obvious. First, the regions of high velocity (dark colour) at the nematode's midsection seem to change and intensify as XG concentration and ST effects are increased. Second, there is an overall decrease in average velocity near the head and an increase near the tail as XG concentration increases. These data show that velocity fields generated by swimming nematodes can be modified by ST effects, even as the imposed swimming kinematics remain unchanged.

\begin{figure}
  \centerline{\includegraphics[width=8.5cm]{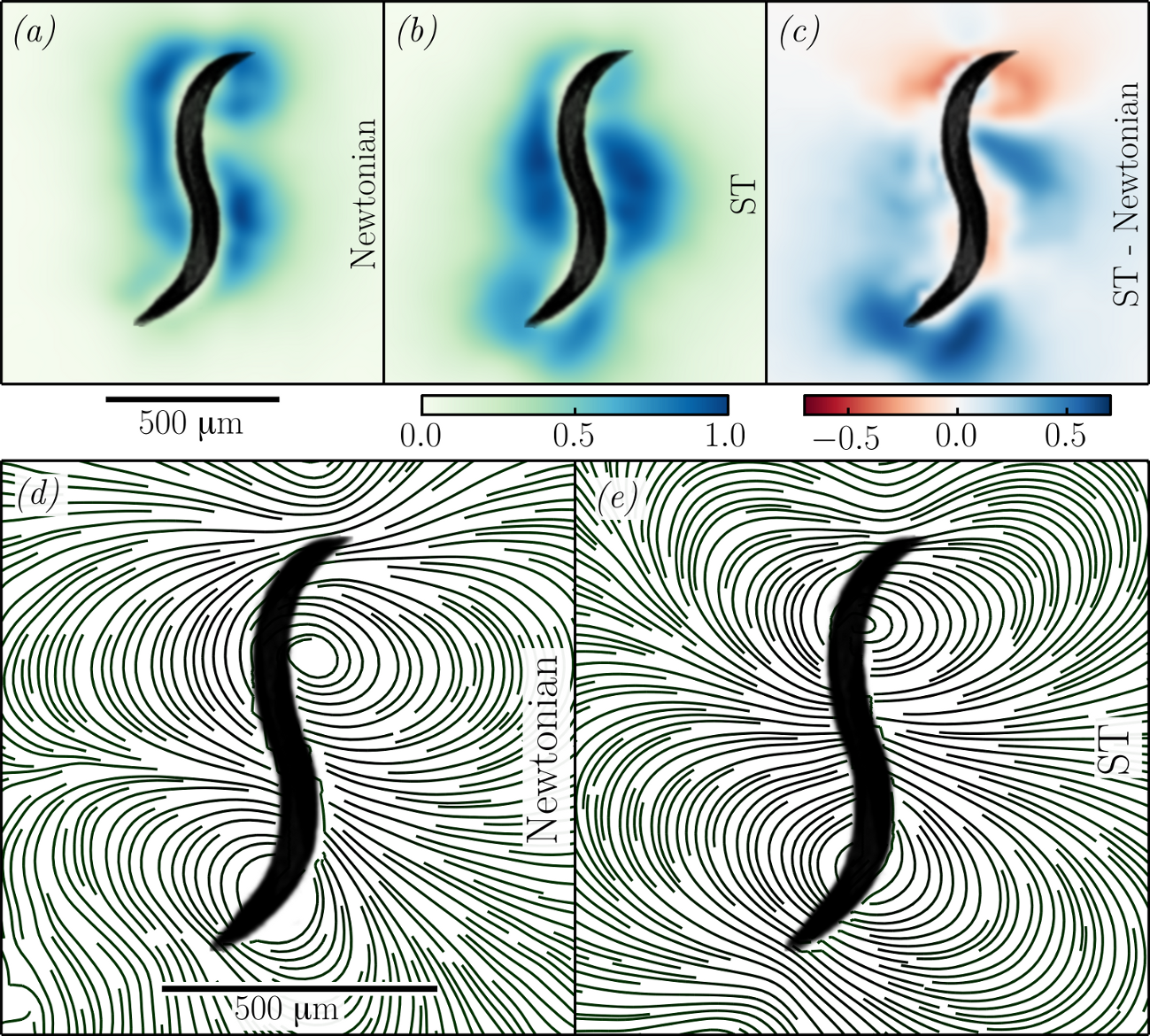}}
  \caption{(Colour available online) \textit{(a-c)} Normalized snapshots of velocity magnitude \textit{(a)} $\hat{v}_N$, for a Newtonian fluid ($\eta = 1$ mPa$\cdot$s, $\lambda_{Cr} = 0$ s, $n = 1$) and \textit{(b)} $\hat{v}_{ST}$ for a ST fluid ($\eta_{eff} \approx 11$ mPa$\cdot$s, $\lambda_{Cr} \approx 7$ s, $n \approx 0.5$).  \textit{(c)} Snapshot of their difference $\Delta \hat{v} = \hat{v}_{ST} - \hat{v}_N$. \textit{(d, e)} Streamlines computed for the same phase for \textit{(d)} Newtonian and  \textit{(e)} ST fluids.}
\label{fig5}
\end{figure}

To better examine shear-thinning effects on the velocity fields, we focus on one particular phase of the beating cycle. While we expect ST effects at every phase, we examine the phase with the largest average shear rate, since this snapshot should reveal the most pronounced differences between Newtonian and ST fluids. Figure~\ref{fig5} shows the velocity magnitude at one phase in \textit{(a)} M9 buffer solution and \textit{(b)} 1000~ppm XG solution. To reduce noise and compensate for differences in bulk viscosity, we normalize the snapshots by their maximum velocity magnitudes: $\hat{v} = |\boldsymbol{v}|/|\boldsymbol{v}|_{max}$; note that $|\boldsymbol{v}|_{max}$ is similar among fluids (figure~\ref{fig4}). Lastly, we subtract the normalized Newtonian field from the normalized ST field; this difference, $\Delta \hat{v} = \hat{v}_{ST} - \hat{v}_N$ is shown in figure~\ref{fig5}\textit{(c)}. We again find an increase in velocity near the tail and a decrease in velocity near the head. 

Figure~\ref{fig5}\textit{(c)} also reveal a stark difference near the midsection of the worm, as highlighted by the positive region that begins just below the nematode's head and extends to the right, away from the swimmer. The Newtonian fluid velocity in this region is nearly zero, but for the ST case, this region is part of a large, high-velocity lobe. This same region is the site of differences in fluid streamlines, (figure \ref{fig5}\textit{(d)} and \textit{(e)}) for the Newtonian and ST cases, respectively. The streamlines suggest that in the ST case, the characteristic ``head'' and ``tail'' vortices have shifted towards the head and appear more concentrated. The difference in velocity magnitudes near the midsection is located near the center of the head vortex, suggesting a potential way to quantify shear-thinning effects.

\section{Quantifying the Role of Shear-Thinning}
So far we have shown that ST viscosity has negligible effect on the nematode's swimming kinematics (figure~\ref{fig3}) but it does seem to affect the velocity fields generated by swimming (figure~\ref{fig5}). Here, we quantify those effects and seek to connect them with the fluids' shear-thinning rheology. We begin by inspecting the flow structure (vortex) near the nematode's head for both the Newtonian and ST fluids (figure~\ref{fig5}\textit{(d)} and \textit{(e)}). For each fluid, we take the velocity field at the same phase as in figure~\ref{fig5}, sampled at a grid of points spaced 21~\si{\micro}m apart. We compute vorticity $\boldsymbol{w}$ in two dimensions, so that $w_z \equiv {\partial v_y}/{\partial x} - {\partial v_x}/{\partial y}$. 

\begin{figure}
  \centerline{\includegraphics[width=10cm]{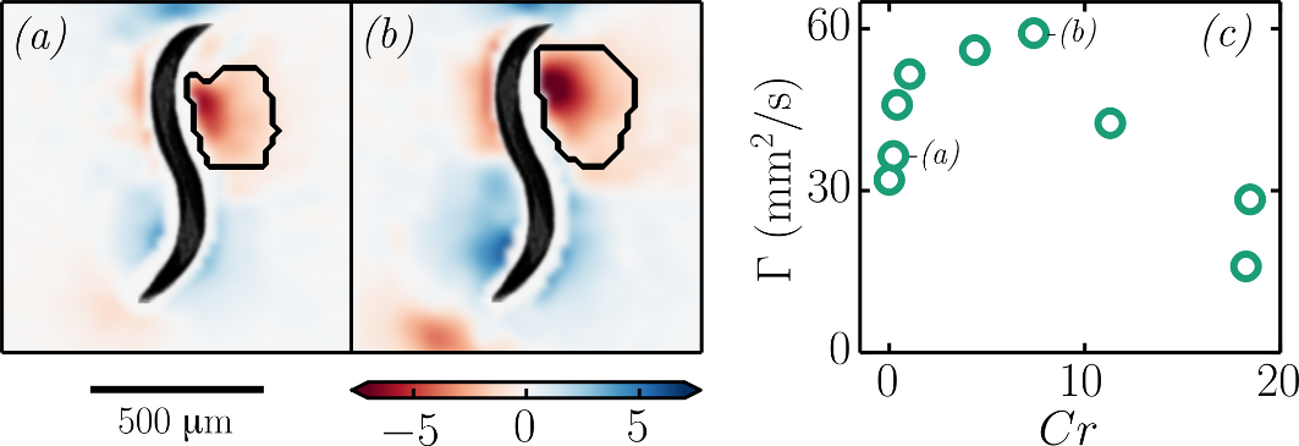}}
  \caption{(Colour available online) Vorticity map calculated for \textit{C.\ elegans} swimming in \textit{(a)} 50 ppm XG in M9 and \textit{(b)} 500 ppm XG in M9. The region of the head vortex (outlined in black) is defined as the region with vorticity greater than 10\% of the local maximum. \textit{(c)} Circulation $\Gamma$ of the head vortex for M9 and ST fluids as a function of the expected local importance of shear thinning, $Cr=\lambda_{Cr}\overline{\dot{\gamma}}$, where $\lambda_{Cr}$ is the Carreau timescale and $\overline{\dot{\gamma}}$ is the measured average local shear rate.}
\label{fig6}
\end{figure}

Figures~\ref{fig6}\textit{(a)} and \textit{(b)} show vorticity fields for the 50 and 500~ppm XG solutions, respectively, with the head vortices outlined; note that the 50 ppm XG vorticity map is not substantially different from the Newtonian case (not shown). We define the head vortex as the region near the head with $|\boldsymbol{w}|$ greater than 10\% of the local maximum. The head vortex is in fact the region of highest vorticity over the entire flow field. The two vorticity fields suggest that increasing the concentration of XG, and thus the ST behaviour of the fluid, increases the magnitude and size of the head vortex. We note that this is not simply due to a change in bulk or ``average'' viscosity, since increasing viscosity would merely scale the vorticity field without changing its shape. 

Next, we measure the circulation or vorticity flux $\Gamma$, which can be thought of as the total vorticity normal to the surface of interest. We compute $\Gamma$ as the discrete integral of vorticity $w_{z,i}$ at each grid cell $i$ within the vortex region: $\Gamma = \sum \, a w_{z,i}$, where $a$ is the area of a grid cell. We also use the velocity field and fluid rheology to estimate the \textit{expected} importance of ST behaviour in the vortex region. This is quantified with a Carreau number $Cr=\lambda_{Cr} \overline{\dot{\gamma}}$. The characteristic vortex shear rate $\overline{\dot{\gamma}}$ is the root-mean-squared average over all grid cells in the vortex, where each cell's local shear rate is computed as $\dot{\gamma} \equiv {\partial v_x}/{\partial y} + {\partial v_y}/{\partial x}$.

Figure~\ref{fig6}\textit{(c)} summarizes our measurements of circulation $\Gamma$ for a range of Carreau numbers $Cr$, corresponding to a Newtonian fluid (M9, $Cr = 0$) and all tested XG concentrations. We find that circulation is generally enhanced as ST effects are increased. We also observe a similar trend in vorticity magnitude both temporally and spatially for the entire cycle. At high $Cr$ (corresponding to high XG concentrations), the increase in overall viscosity appears to suppress circulation: these data points are close to the power-limited regime, and for fluids above $Cr \approx 15$, the shear stresses needed to access the ST regime exceed $\sigma = 0.1$~Pa. However, for $0<Cr<15$ (XG concentrations $\leq 1000$~ppm), the measured circulation and vorticity are larger than in the Newtonian case, and this suggests agreement with previous theoretical work, which predicted enhanced vorticity in ST fluids~\citep{Velez2013}.

Lastly, we would like to quantify the decrease in fluid velocity near the head and the accompanying increase in fluid velocity near the tail observed in Figs.~\ref{fig4} and~\ref{fig5}. In order to measure these effects, we divide the flow field region into three sections: ``head,'' ``mid,'' and ``tail.'' The head-section is defined as the region of the $y$-axis which contains the first 30\% of the worm body ($0<s<0.3L$) (figure~\ref{fig1}\textit{(a)}), the tail-section is defined as the region which contains the last 30\% of the worm body ($0.7L<s<L$), and each section extends $0.5L$ in the positive and negative $x$-directions. 

\begin{figure}
  \centerline{\includegraphics[width=9.5cm]{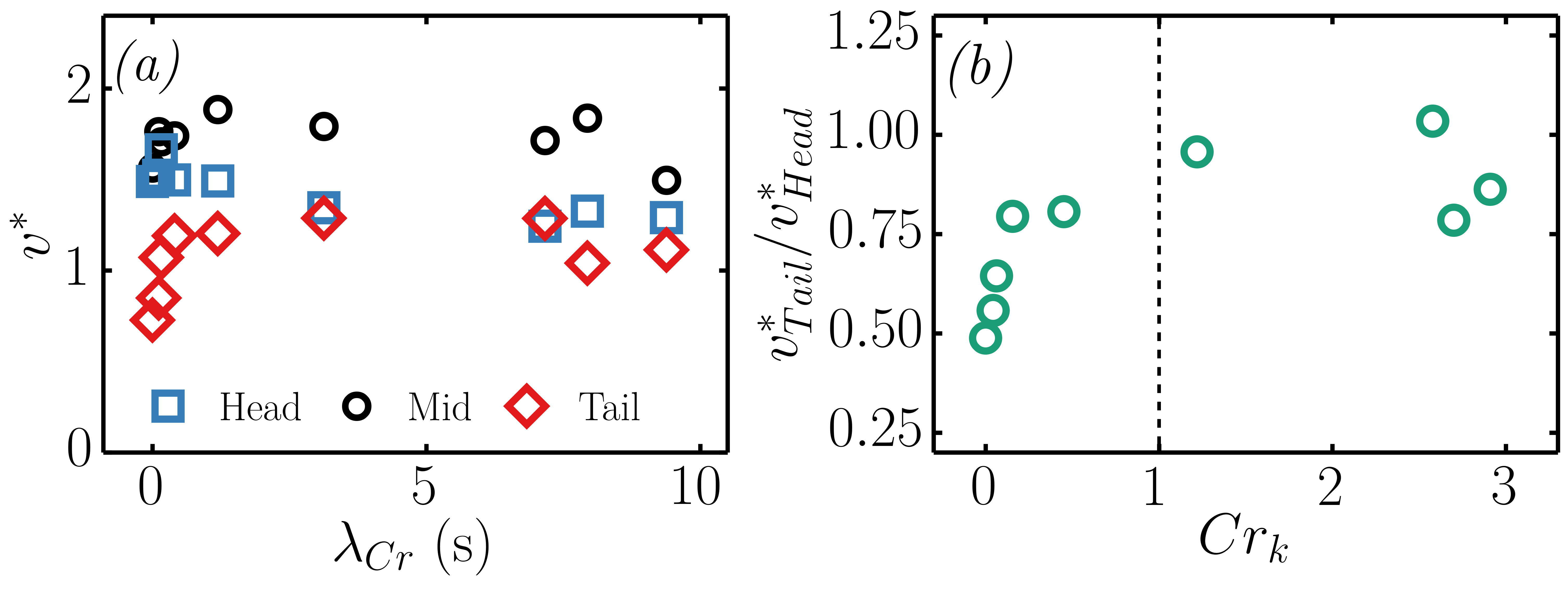}}
  \caption{(Colour available online) Local velocity analysis of the increase in fluid flow near the tail of the nematode. \textit{(a)} Spatially averaged, RMS-normalized fluid velocity near the head ({\scriptsize \color{blue}$\bm{\square}$}), midsection ($\bm{\circ}$), and tail ({\color{red}$\bm{\diamond}$}) of the nematode as a function of Carreau timescale $\lambda_{Cr}$. \textit{(b)} The ratio of the fluid velocity at the tail to fluid velocity at the head as a function of the kinematic Carreau number $Cr_k = \lambda_{Cr}(U/L)$.}
\label{fig7}
\end{figure}

Figure~\ref{fig7}\textit{(a)} shows the normalized average velocity magnitude $v^*$ in each region as a function of the Carreau timescale $\lambda_{Cr}$. We define $v^*$ as $\left\langle v \right\rangle / \sqrt{\left\langle v^2 \right\rangle}$, where $\left\langle v \right\rangle$ is the \textit{local} average fluid velocity in each region and $\sqrt{\left\langle v^2 \right\rangle}$ is the \textit{global} root-mean-squared velocity. Increasing $\lambda_{Cr}$ corresponds to increasing XG concentration, with $\lambda_{Cr} = 0$ for the M9 buffer. We find that \textit{(i)} $v_{Head}^*$ decreases slightly with increasing $\lambda_{Cr}$, \textit{(ii)} $v_{Tail}^*$ increases dramatically with increasing $\lambda_{Cr}$ and plateaus at $\lambda_{Cr} \approx 3$~s, and \textit{(iii)} $v_{Mid}^*$ is largely insensitive to increasing $\lambda_{Cr}$. 

To understand why shear-thinning viscosity might affect the velocity field at the head and tail differently, we consider an element of ST fluid as it passes from head to tail of the forward-swimming nematode. When the shear strain rate of this element changes, the characteristic time for it to approach its new steady-state viscosity is $\lambda_{Cr}$~\citep{Carreau1997}. We therefore define a \emph{kinematic} Carreau number, $Cr_k = \lambda_{Cr} (U / L)$, where $U$ and $L$ are the nematode's speed and length, in order to compare $\lambda_{Cr}$ with the characteristic time for fluid to pass from head to tail. If $Cr_k \gtrsim 1$, fluid passing the tail will still be thinned from its earlier encounter with the moving head, in addition to the local, instantaneous body motion. However, if $Cr_k \ll 1$, viscosity near the tail will be a function solely of the local, instantaneous shear rate.

This phenomenon is made more apparent by plotting the ratio $v_{Tail}^*/v_{Head}^*$ as a function of $Cr_k$. Figure~\ref{fig7}\textit{(b)} shows that this ratio increases monotonically with $Cr_k$, and becomes asymptotic near $Cr_k = 1$. These observations suggest that the velocity of fluid near the tail is significantly enhanced when that fluid is under the influence of prior shear-thinning. Also, the variation in $v_{Tail}^*/v_{Head}^*$ at $Cr_k \approx 3$ suggests that once again we are nearing the nematode's power-limited regime, in which the bulk viscosity seems to overwhelm ST effects. This effect of the persistence of shear-thinning, at the scale of the entire body, is qualitatively consistent with previous numerical results showing a large envelope of shear-thinned fluid along the body of a uniflagellated swimmer~\citep{MJ2013}. 

\section{Conclusion}
We have investigated the effects of ST viscosity on the swimming kinematics and flow fields of the nematode \textit{C.\ elegans} in solutions of a semi-rigid, rod-like polymer (XG) in buffer solution (M9). We find no significant differences between the observed kinematics in ST fluids and the kinematics in Newtonian fluids of similar viscosity, a result consistent with recent theoretical calculations~\citep{Velez2013}. Despite this experimental observation (unmodified kinematics), we find substantial differences between the resulting Newtonian and ST flow fields. 

Temporal and spatial averages of velocity fields, along with a snapshot of the flow fields at one phase in the cycle, reveal a structural difference in the flow field near the head of the worm in ST fluids when compared to the Newtonian case. Streamlines reveal the major structural difference occurs within the dominant body vortex near the nematode's head; we quantify this effect by computing the circulation of the vortex, thereby accounting for strength and size. We find that an increase in shear-thinning behaviour leads to an enhancement in circulation, consistent with recent theoretical calculations (see \citet{Velez2013}). Furthermore, the ST flow fields reveal a decrease in average fluid velocity near the head of the organism and an increase near the tail. We find that the ratio of average fluid velocity at the tail to that at the head increases as the Carreau timescale $\lambda_{Cr}$ increases, and that this effect plateaus once $\lambda_{Cr}$ is similar to the timescale of the forward swimming motion, $U/L$.

Numerous biological systems, such as beating cilia, spermatozoa, and bacteria, experience natural environments composed of non-Newtonian, shear-thinning fluids. This experimental study systematically explores the effects of shear-thinning viscosity on the kinematics and flow fields of an model undulatory swimmer. We therefore believe this work brings us another step closer to understanding swimming in complex fluids. 

\section*{Acknowledgements}
We thank T. Lamitina and X. Shen for help with experiments. This work was supported by NSF-CAREER-0954084 and by the Penn NSF MRSEC (DMR-1120901). 

\bibliographystyle{jfm}
\bibliography{shearThinning}

\begin{thebibliography}{25}
\expandafter\ifx\csname natexlab\endcsname\relax\def\natexlab#1{#1}\fi

\bibitem[Alexander(1991)]{Alexander1991}
{\sc Alexander, M.} 1991 {\em Introduction to soil microbiology\/}. Malabar,
  FL: R.E. Krieger.

\bibitem[Brenner(1974)]{Brenner1974}
{\sc Brenner, S.} 1974 The genetics of \emph{Caenorhabditis elegans}. {\em
  Genetics\/} {\bf 77}, 71--94.

\bibitem[Carreau {\em et~al.\/}(1997)Carreau, DeKee \& Chhabra]{Carreau1997}
{\sc Carreau, P.J., DeKee, D.C.R. \& Chhabra, R.P.} 1997 {\em Rheology of
  Polymeric Systems\/}. Munich: Hanser.

\bibitem[Dasgupta {\em et~al.\/}(2013)Dasgupta, Liu, Fu, Berhanu, Breuer,
  Powers \& Kudrolli]{Dasgupta2013}
{\sc Dasgupta, M., Liu, B., Fu, H.C., Berhanu, M., Breuer, K.S., Powers, T.R.
  \& Kudrolli, A.} 2013 Speed of a swimming sheet in newtonian and viscoelastic
  fluids. {\em Phys. Rev. E\/} {\bf 87}, 013015.

\bibitem[Fauci \& Dillon(2006)]{Fauci2006}
{\sc Fauci, L.J. \& Dillon, R.} 2006 Biofluidmechanics of reproduction. {\em
  Annu. Rev. Fluid Mech.\/} {\bf 38}, 371--394.

\bibitem[Fu {\em et~al.\/}(2010)Fu, Shenoy \& Powers]{Fu2010}
{\sc Fu, H.C., Shenoy, V.B. \& Powers, T.R.} 2010 Low-{R}eynolds-number
  swimming in gels. {\em EPL\/} {\bf 91}.

\bibitem[Fu {\em et~al.\/}(2009)Fu, Wolgemuth \& Powers]{Fu2009}
{\sc Fu, H.C., Wolgemuth, C.W. \& Powers, T.R.} 2009 Swimming speeds of
  filaments in nonlinearly viscoelastic fluids. {\em Phys. Fluids\/} {\bf 21},
  033102--033110.

\bibitem[Gagnon {\em et~al.\/}(2013)Gagnon, Shen \& Arratia]{Gagnon2013}
{\sc Gagnon, D.A., Shen, X.N. \& Arratia, P.E.} 2013 Undulatory swimming in
  fluids with polymer networks. {\em Europhys. Lett.\/} {\bf 104}, 14004.

\bibitem[Guasto {\em et~al.\/}(2010)Guasto, Johnson \& Gollub]{Guasto2010}
{\sc Guasto, J.S., Johnson, K.A \& Gollub, J.P.} 2010 Oscillatory flows induced
  by microorganisms swimming in two dimensions. {\em Phys. Rev. Lett.\/} {\bf
  105}, 168102.

\bibitem[Harman {\em et~al.\/}(2012)Harman, Dunham-Ems, Caimano, Belperron,
  Bockenstedt, Fu, Radolf \& Wolgemuth]{Harman2012}
{\sc Harman, M.W., Dunham-Ems, S.M., Caimano, M.J., Belperron, A.A.,
  Bockenstedt, L.K., Fu, H.C., Radolf, J.D. \& Wolgemuth, C.W.} 2012 The
  heterogenous motility of the {L}yme disease spirochete in gelatin mimics
  dissemination through tissue. {\em Proc. Natl. Acad. Sci. USA\/} {\bf 109},
  3059--3064.

\bibitem[Lauga(2007)]{Lauga2007}
{\sc Lauga, E.} 2007 Propulsion in a viscoelastic fluid. {\em Phys. Fluids\/}
  {\bf 19}, 083104--083113.

\bibitem[Lauga \& Goldstein(2012)]{Lauga2012}
{\sc Lauga, E. \& Goldstein, R.E.} 2012 Dance of the microswimmers. {\em Phys.
  Today\/} {\bf 65}, 30--35.

\bibitem[Lauga \& Powers(2009)]{Lauga2009}
{\sc Lauga, E. \& Powers, T.R.} 2009 The hydrodynamics of swimming
  microorganisms. {\em Reports on Progress in Physics\/} {\bf 72}, 096601.

\bibitem[Liu {\em et~al.\/}(2011)Liu, Powers \& Breuer]{Liu2011}
{\sc Liu, B., Powers, T.R. \& Breuer, K.S.} 2011 Force-free swimming of a model
  helical flagellum in viscoelastic fluids. {\em Proc. Natl. Acad. Sci. USA\/}
  {\bf 108}, 19516--19520.

\bibitem[Montenegro-Johnson {\em et~al.\/}(2012)Montenegro-Johnson, Smith,
  Smith, Loghin \& Blake]{MJ2012}
{\sc Montenegro-Johnson, T.D., Smith, A.A., Smith, D.J., Loghin, D. \& Blake,
  J.R.} 2012 Modelling the fluid mechanics of cilia and flagella in
  reproduction and development. {\em Eur. Phys. J. E\/} {\bf 35}, 111.

\bibitem[Montenegro-Johnson {\em et~al.\/}(2013)Montenegro-Johnson, Smith \&
  Loghin]{MJ2013}
{\sc Montenegro-Johnson, T.D., Smith, D.J. \& Loghin, D.} 2013 Physics of
  rheologically enhanced propulsion: {D}ifferent strokes in generalized
  {S}tokes. {\em Phys. Fluids\/} {\bf 25}, 081903.

\bibitem[Purcell(1977)]{Purcell1977}
{\sc Purcell, E.M.} 1977 Life at low {R}eynolds number. {\em Am. J. Phys.\/}
  {\bf 45}~(1), 3--11.

\bibitem[Rankin(2002)]{Rankin2002}
{\sc Rankin, C.H.} 2002 From gene to identified neuron behavior in
  \textit{{C}aenorhabditis elegans}. {\em Nat. Rev. Genet.\/} {\bf 3},
  622--630.

\bibitem[Saintillan \& Shelley(2012)]{Saintillan2012}
{\sc Saintillan, D. \& Shelley, M.J.} 2012 Emergence of coherent structures and
  large-scale flows in motile suspensions. {\em J. R. Soc. Interface\/} {\bf
  9}, 571--585.

\bibitem[Shen \& Arratia(2011)]{Shen2011}
{\sc Shen, X.N. \& Arratia, P.E.} 2011 Undulatory swimming in viscoelastic
  fluids. {\em Phys. Rev. Lett.\/} {\bf 106}, 208101.

\bibitem[Sznitman {\em et~al.\/}(2010{\natexlab{{\em a\/}}})Sznitman, Shen,
  Sznitman \& Arratia]{Sznitman2010PoF}
{\sc Sznitman, J., Shen, X.N., Sznitman, R. \& Arratia, P.E.}
  2010{\natexlab{{\em a\/}}} Propulsive force measurements and flow behavior of
  undulatory swimmers at low {R}eynolds number. {\em Phys. Fluids\/} {\bf 22},
  121901.

\bibitem[Sznitman {\em et~al.\/}(2010{\natexlab{{\em b\/}}})Sznitman, Gupta,
  Hager, Arratia \& Sznitman]{RSznitman2010}
{\sc Sznitman, R., Gupta, M., Hager, G.D., Arratia, P.E. \& Sznitman, J.}
  2010{\natexlab{{\em b\/}}} Multi-environment model estimation for motility
  analysis of \textit{{C}aenorhabditis elegans}. {\em PLoS ONE\/} {\bf 5},
  e11631.

\bibitem[Teran {\em et~al.\/}(2010)Teran, Fauci \& Shelley]{Teran2010}
{\sc Teran, J., Fauci, L. \& Shelley, M.} 2010 Viscoelastic fluid response can
  increase the speed and efficiency of a free swimmer. {\em Phys. Rev. Lett.\/}
  {\bf 104}, 038101.

\bibitem[V\'{e}lez-Cordero \& Lauga(2013)]{Velez2013}
{\sc V\'{e}lez-Cordero, J.N. \& Lauga, E.} 2013 Waving transport and propulsion
  in a generalized {N}ewtonian fluid. {\em J. Non-Newton. Fluid.\/} {\bf 199},
  37--50.

\bibitem[Wyatt \& Liberatore(2009)]{Wyatt2009}
{\sc Wyatt, N.B. \& Liberatore, M.W.} 2009 Rheology and viscosity scaling of
  the polyelectrolyte xanthan gum. {\em J. Appl. Polym. Sci.\/} {\bf 114},
  4076--4084.

\end{thebibliography}

\end{document}